\documentclass[rnote]{aa}  
\usepackage{graphicx}
\usepackage{txfonts}
\usepackage{natbib}
\usepackage{longtable}
\begin{document}
	\title{30~GHz flux density measurements of the Caltech-Jodrell flat-spectrum sources with OCRA-p}

	\subtitle{}

	\author{S. R. Lowe
		\inst{1}
		\and
		M. P. Gawro\'{n}ski\inst{2}
		\and 
		P. N. Wilkinson\inst{1}
		\and
		A. J. Kus\inst{2}
		\and
		I. W. A. Browne\inst{1}
		\and
		E. Pazderski\inst{2}
		\and
		R. Feiler\inst{2}
		\and
		D. Kettle\inst{1}
	}

	\offprints{S. R. Lowe}

	\institute{
		University of Manchester, Jodrell Bank Observatory, Macclesfield, Cheshire, SK11 9DL, UK \\
		\email{Stuart.Lowe@manchester.ac.uk}
	\and
		Toru\'{n} Centre for Astronomy, Nicolaus Copernicus University, 87-148 Toru\'{n}/Piwnice, Poland
	}

   \date{Received June 7, 2007; accepted ???}

 
	\abstract
	{}
	{To measure the 30-GHz flux densities of the $293$ sources in the Caltech-Jodrell Bank flat-spectrum (CJF) sample. The measurements are part of an ongoing programme to measure the spectral energy distributions of flat spectrum radio sources and to correlate them with the milliarcsecond structures from VLBI and other measured astrophysical properties.}
	{The 30-GHz data were obtained with a twin-beam differencing radiometer system mounted on the Toru\'{n} 32-m telescope. The system has an angular resolution of $1.2$\arcmin.}
	{Together with radio spectral data obtained from the literature, the 30-GHz data have enabled us to identify $42$ of the CJF sources as Giga-hertz Peaked Spectrum (GPS) sources. Seventeen percent of the sources have rising spectra ($\alpha > 0$) between $5$ and $30$~GHz.}
	{}


   \maketitle
%

\section{Introduction}

	The emission from most flat-spectrum radio sources, from radio frequencies through gamma-rays, is thought to arise in relativistic jets and be beamed synchrotron self-Compton emission.  Often described as blazar emission it is characterized by two peaks in the spectral energy distribution (SED), one synchrotron and one inverse Compton. From object to object the peak frequency can occur anywhere between $10^{10}$~Hz to $10^{15}$~Hz. There are claims that where the peaks occur depends systematically on radio luminosity \citep[][]{1998MNRAS.299..433F,2002A&A...386..833G}. The correlation is in the sense that the synchrotron peak frequency increases as the luminosity decreases. This is potentially an important result but has been questioned by several authors \citep[e.g.][]{2005MNRAS.356..225A}. A major problem is, however, the lack of good quality SEDs on well-defined samples of objects and for this reason we have embarked on a programme to try to rectify this deficiency. Flux density measurements at centimeter wavelengths and shorter are the most important in order to define the position of the synchrotron peak. Here we report measurements with a new receiver, OCRA-p on the Torun 32m Telescope at a wavelength of 1cm.   

	The CJF sample is currently the best studied sample of flat-spectrum radio sources. The CJF sample \citep{CJF} consists of $293$ sources selected from three previous VLBI surveys: the PR survey \citep{PR}, the first Caltech-Jodrell (CJ) survey \citep[CJ1:][]{CJ1} and the second Caltech-Jodrell survey \citep[CJ2:][]{1994ApJS...95..345T}. The selection criteria were:
	\begin{enumerate}\setlength{\itemsep}{2mm}
	\item $S_{4.85~GHz} \ge 350~mJy$
	\item $\alpha^{4.85~GHz}_{1.4~GHz} \ge -0.5$\footnote{$S\propto\nu^{\alpha}$}
	\item $\delta(1950) \ge 35^\circ$
	\item $|b| \ge 10^\circ$
	\end{enumerate}
	In addition to the structural information  obtained in the CJ VLBI surveys, extensive follow-up observations have been made with the VLBA (Britzen et al. in prep) to study the statistics of superluminal motions; redshift information is available for $>90\%$ of the sample. Furthermore, all $293$ sources have been observed in soft X-rays, either as part of the \emph{ROSAT} All-Sky Survey or in \emph{ROSAT} pointed observations \citep{2002evn..conf...99B}.  The CJF sample is therefore a natural starting point for a programme aimed at understanding the relationships between the broad-band SEDs and the spatial structure, kinematics and X-ray properties of compact radio sources. Several different types of objects are found in samples selected, like CJF, from radio surveys made at relatively low (few GHz) frequencies. While the sample is dominated by highly-relativistic ``core-jet'' sources it also contains the precursors of  large-scale double sources \citep[the Compact Symmetric Objects, CSOs: e.g.][]{1994ApJ...432L..87W,1996ApJ...460..634R,1996ApJ...460..612R} and perhaps a few ULIRGS (typically more distant and luminous versions of the starburst galaxy, Arp 220). VLBI observations immediately distinguish between these different types: the core-jet sources are highly asymmetric, in contrast to the CSOs, while ULIRGS are highly resolved.

	The complementary radio spectral information is available from hundreds of MHz to typically $\sim10$~GHz but information at higher frequencies is sparse. As part of the commissioning programme for the new One Centimetre Radio Array prototype (OCRA-p) we have measured the flux densities of all the CJF sources at $30$~GHz using the Toru\'{n} 32-m Telescope. In addition to their intrinsic astrophysical interest the CJF sources are relatively strong and hence provide a first test of the OCRA-p system. Other early results with the OCRA-p system are presented elsewhere \citep{Lancaster2007}.

	In sections 2 and 3 we briefly describe the OCRA-p system and the observing techniques we used to take the 30-GHz data. In section 4 we present the results in the form of a Table of 30-GHz flux densities. In section 5 we discuss the contribution of the new data to the classification of the CJF sample into source types, in particular CSO/GPS sources, and count the number of sources whose spectra are rising at $30$~GHz.

\section{OCRA-p on the Toru\'{n} 32-m telescope}
\label{ocrap}

	The OCRA-p receiver was constructed by the University of Manchester. The basic design of the radiometer was taken from the prototype demonstrator radiometer for the Planck Low Frequency Instrument \citep[LFI,][]{2000ApL&C..37..151M} and is similar to the K-band receivers onboard the WMAP satellite \citep{2003ApJS..145..413J}.

	\begin{table}
	\caption[Nominal system specifications.]{\label{specs} Nominal system specifications for OCRA-p on the Toru\'{n} 32-m telescope.}
	\begin{center}
	\begin{tabular}{ r c }
	\hline
	Frequency Range & $27 - 33$~GHz	\\
	System Temperature, $T_{sys}$ & $40$~K	\\
	Individual beam FWHM	& $1.2$\arcmin	\\
	Beam spacing	& $3.1$\arcmin	\\
	Aperture Efficiency	& $45\%$	\\ \hline
	\end{tabular}
	\end{center}
	\end{table}
	
	Rather than switching between a source and a fixed temperature load, as in Planck, OCRA-p switches between two closely spaced horns observing the sky continuously. When mounted on the Toru\'{n} 32-m, the two beams are separated by $3.1$\arcmin, a figure set by the optics of the telescope, and therefore both receive very similar atmospheric contributions from the near-field. A phase switch in one of the radiometer arms alternates the input signals between the two output channels and, by taking the difference between the channels and between switch states, it is possible to reduce the effects of both the atmospheric and gain fluctuations in the back-end amplifiers. The differencing scheme produces positive and negative beams (see Figure \ref{fig:scanstrategy}) and this is characterised by an s-shaped response. The nominal specifications for OCRA-p are given in Table \ref{specs}. The construction and commissioning of the instrument is described in detail in \cite{Lowe2005}.

	The switching scheme works well despite the low altitude ($200$~m) of the Toru'{n} site. OCRA-p has been able to make good observations even during summer nights. 

\section{Observations}
\label{obs}

\begin{figure}
	\includegraphics[angle=0,scale=.44]{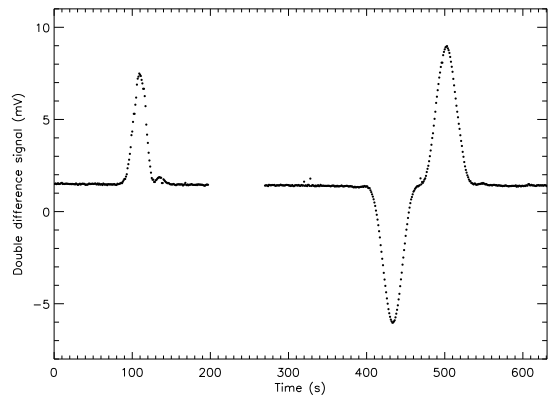}
	\caption{A figure to illustrate the scan strategy. An initial pointing correction is found by fitting a Gaussian function to an elevation scan (left) through the source. This correction is applied to the telescope position and then an azimuthal scan is performed. The initial pointing error can be seen in the difference in amplitudes of the beams in the two scans. The elevation scan did not pass through the centre of the source, so the amplitude is lower than the azimuth scan (right) with the elevation correction. }
	\label{fig:scanstrategy}
	\end{figure}

	At the time these observations were taken {\bf(April - August 2005)}, the pointing accuracy of the 32-m radio telescope was dependent on elevation and azimuth in a complex manner. The pointing accuracy was best at relatively high elevations ($50^\circ - 70^\circ$) and so it was desirable to confine most of our observations to this range. A pointing observation was incorporated within the overall observing strategy for each source. For each source measurement, an elevation scan of the source was first recorded. Custom software then fitted a Gaussian function to these data, in real time, and the central point was used to generate a pointing correction for the elevation coordinate. With the calculated elevation correction an azimuthal scan was then performed. The length of the scan was chosen to be several beam spacings in extent to allow for azimuthal pointing errors and to determine a good baseline. Figure \ref{fig:scanstrategy} shows an indicative illustration of the scan strategy. To the left is the preliminary elevation scan and to the right, the azimuthal scan. The s-shaped response, generated as the source passes through each beam in turn, can clearly be seen in the azimuthal scan.

	For each azimuthal scan a baseline and two independent Gaussian functions were fitted for each beam. At this stage, every observation that was significantly affected by atmospheric fluctuations, or that showed large asymmetries in the amplitudes of the beams, was rejected. 

	The amplitudes of the fitted Gaussian functions are directly related to the strength of the source. To convert the measured signal to a flux density it is necessary to calibrate against a source of known strength. Our primary flux calibrator was chosen to be the planetary nebula NGC\,7027 which has an optical diameter of $\sim15$\arcsec and is circumpolar at the latitude of Toru'{n}. There is a recent absolute temperature calibration of NGC\,7027 performed by \cite{1999AJ....118.2908M} who report a flux density of $5.45\pm0.20$~Jy at $32$~GHz. Assuming a spectral index of $-0.1\pm0.1$ and estimating the beam dilution factor, using a simple geometric model of the emission created from 6-cm radio maps from \citet[][]{1997MNRAS.284..815B}, gives an expected flux density of $5.46\pm0.20$~Jy at $30$~GHz. All the final flux density values have been scaled to this figure.

	Although NGC\,7027 is circumpolar, it is not always at a suitable elevation and can be too far from the CJF target sources for reliable calibration. A network of secondary flux density calibrators was therefore generated. The secondary calibrators were: 0010+405, 0016+731, 0109+351, 0309+411, 0633+734, 0642+449, 0821+394, 0833+585, 0955+476, 1044+719, 1144+402, 1418+546, 1638+398, 1758+388, 2007+777, 2021+614. Several of these sources show long-term variability in their emission, however the level of intrinsic intra-day variations did not contribute significantly to our final error budget.

	Every three to six target observations, a secondary calibrator source was observed. The secondary calibrator measurements were calibrated using the measurements of NGC\,7027. {\bf The final flux densities take into account atmospheric absorption (estimated uncertainty $\sim2\%$ in the correction) and a gain-elevation correction ($\sim2.5\%$). Uncertainties in these corrections as well as pointing errors ($<1\%$ for this observing method), uncertainties in the extrapolated flux density of NGC7027 ($3.6\%$) and in the secondary calibration ($\sim6\%$) have been incorporated into the final error budget. The typical uncertainty for a single flux density measurement is in the range $8 - 10\%$. Since six or more measurements were usually made for each source and the errors are independent, the final uncertainties are typically around $5\%$.}
	
	{\bf The flux density measurements are made over a $6$~GHz bandwidth ($27 - 33$~GHz). We note that sources with spectral curvature that is different to the primary calibrator will have systematic errors in their flux densities. However, as NGC\,7027 and the majority of our sample are flat-spectrum, we neglect this effect given the other uncertainties.}
	
\section{Results}
\label{res}

	The final 30-GHz flux densities are given in Table \ref{tbl:fluxes}. Column (1) gives the IAU B1950 source name for consistency with previous CJF observations by other authors. The J2000 coordinates are given in Columns (2) and (3) and are taken from \cite{CJF}. The flux densities at $5$~GHz, Column (4), are also taken from \cite{CJF} and are reproduced for comparison with those measured by OCRA-p at $30$~GHz in Column (5). Column (6) lists the associated uncertainties in the $30$~GHz flux densities. The spectral index between $1.4$~GHz and $5$~GHz, $\alpha_{1.4}^{5}$, from \cite{CJF} is given in Column (7) and the spectral index between $5$~GHz and $30$~GHz, $\alpha_{5}^{30}$, is given in Column (8). {\bf The formal uncertainty on $\alpha_{5}^{30}$, given the typical uncertainties in both the $5$ and $30$~GHz measurements, is $<0.1$. We note however, that the uncertainties for a few objects may be larger due to variability and the fact that the observations at $5$ and $30$~GHz are not co-eval.} The morphology, from inspection of VLBI maps, is given in Column (9). 
	
	To classify the spectral behaviour of the CJF sources, the latest spectral information was obtained from NED. For the spectral plots including the $30$~GHz values, refer to http://www.manchester.ac.uk/jodrellbank/research/ocra/cjf/ 

	\addtocounter{table}{1}

\section{Discussion}
\label{dis}
	
\begin{figure}
	\includegraphics[angle=0,scale=.44]{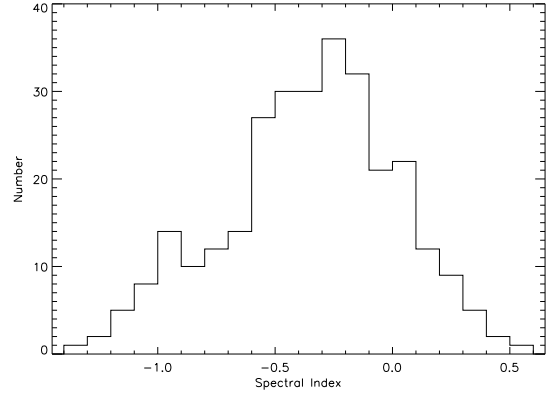}
	\caption{ Distribution of spectral indices calculated between 5~GHz and 30~GHz. Using data from Table \ref{tbl:fluxes}. }
	\label{fig:spectralindex}
	\end{figure}

	Figure \ref{fig:spectralindex} shows the distribution of spectral indices as given in Table \ref{tbl:fluxes}. The distribution can be compared with Figure 1b in \cite{CJF} which shows the spectral index distribution between $1.4$~GHz and $4.85$~GHz, $\alpha_{1.4}^{4.85}$. The major difference is the shift to lower spectral indices mainly due to the spectral index cut-off of the CJF sample ($-0.5$ at $5$~GHz). A significant fraction ($\sim31\%$) of sources are seen to have spectral indices, $\alpha_{5}^{30}$ steeper than $-0.5$. This is the expected behaviour of GPS sources peaking at or above $\sim5$~GHz. Looking at the spectra as a whole we classify $42$ ($\sim14\%$) sources as having GPS spectra (0018+729; 0102+480; 0108+388; 0145+386; 0248+430; 0615+820; 0627+532; 0636+680; 0646+600; 0700+470; 0710+439; 0711+356; 0740+768; 0859+681; 0900+520; 0950+748; 1031+567; 1107+607; 1144+352; 1205+544; 1206+415; 1223+395; 1226+373; 1312+533; 1333+459; 1337+637; 1342+663; 1355+441; 1356+478; 1413+373; 1442+637; 1526+670; 1534+501; 1543+480; 1550+582; 1622+665; 1755+578; 1758+388; 1826+796; 1946+708; 2352+495; 2356+385). We note that \cite{1998PASP..110..493O} found that $\sim10\%$ of sources, selected at $5$~GHz, were GPS. Our results on a $5$~GHz sample, selected against sources with steep spectra, are therefore in rough accord with \cite{1998PASP..110..493O} statistics.
	
	There are a total of $49$ sources ($\sim17\%$) which have $\alpha_{5}^{30} > 0$ and can therefore be considered to have rising spectra. Only one source, 0109+351, has $\alpha_{5}^{30} >0.5$ compared to the 41 which have $\alpha_{1.4}^{5} >0.5$ \citep[][]{CJF}. We do not, however, claim that this is clear evidence for a decrease in the number of steeply rising sources with increasing frequency but is, we suspect, simply a result of selection bias. {\bf If for example we had selected sources at $30$~GHz we assert that we would have found many more sources with spectra steeply rising between $5$ and $30$~GHz.} The same effect is seen in the results reported by \cite{2004MNRAS.354..485B} on the $\alpha_{15}^{43}$ spectral index distribution, for sources selected at 15~GHz from the 9C survey; these indicate that around 1\% of these sources have spectra rising faster than 0.5 between $15$~GHz and $43$~GHz.

	The number of sources with steeply rising spectra in high frequency surveys is of great interest because sources act as a confusing foreground to CMB studies and to SZ decrement measurements. \cite{2003MNRAS.342..915W} have made surveys of areas of sky at 15~GHz coincident with small area CMB fields studied with the Very Small Array \citep[VSA,][]{2003MNRAS.341.1076S}. OCRA-p measurements of these sources will be reported elsewhere. However, with the higher sensitivity over the whole sky which will be achieved by Planck, wide-field 30~GHz surveys are urgently required. We plan to install a 30~GHz 16-beam array (OCRA-F) on the Toru\'{n} telescope in 2008 and one of its main tasks will be to do such a blind survey.

	Further work on spectral energy distributions is in progress and is planned for the future. Observations of 145 (130 have signal-to-noise greater than 3) of the CJF sources have been made at 850~$\mu m$ with SCUBA and will be reported in Anton et al. (2007). With an improved telescope control system it will be possible to improve greatly the speed and accuracy of flux density measurements made with OCRA-p. This will facilitate a much larger programme of blazar flux density measurements and variability monitoring to be undertaken in support of the GLAST satellite which will be launched later this year. 

\begin{acknowledgements}
	We gratefully acknowledge the financial support of the Royal Society Paul Instrument Fund which allowed us to construct the 30~GHz receiver. We are also grateful to the Polish Ministry of Science and Higher Education (grant KBN 5 P03D 024 21) which provided support for operating the receiver on the Toru\'{n} 32-m telescope. This research has made use of the NASA/IPAC Extragalactic Database (NED) which is operated by the Jet Propulsion Laboratory, California Institute of Technology, under contract with the National Aeronautics and Space Administration. S.R Lowe was supported by a PPARC studentship.
\end{acknowledgements}

\bibliographystyle{aa} 
\bibliography{aarefs} 

\longtab{2}{
	\begin{longtable}{c c c c c c c c c}
	\caption{\label{tbl:fluxes} 30-GHz flux density measurements with OCRA-p and the Toru'{n} 32-m telescope.; col. (1) gives the IAU B1950 source name; cols. (2) and (3) give the J2000 positions; col. (4) gives the flux densities at 5~GHz; col. (5) gives the 30-GHz OCRA-p flux densities; col. (6) gives the associated uncertainty; cols. (7) and (8) give the calculated spectral indices between 1.4 and 5~GHz \citep[$\alpha_{1.4}^{5}$ taken from ][]{CJF}, and between 5 and 30~GHz ($\alpha_{5}^{30}$); col. (9) indicates the VLBI morphology where c=compact, cj=core-jet, d=double, cso=compact symmetric object, mso=medium symmetric object, t=triple, g=gravitational lens. } \\
	\hline\hline
	Name	& R.A. (J2000)	& Declination (J2000)	& $S_{5.0}$ (Jy)	& $S_{30}$ (Jy)	& $\Delta S_{30}$ (Jy) & $\alpha^{5}_{1.4}$ & $\alpha^{30}_{5}$	& Morph.	\\
	(1)	& (2)	& (3)	& (4)	& (5)	& (6)	& (7)	& (8)	& (9)	\\
	\hline
	\endfirsthead
	\caption{continued.}\\
	\hline\hline
	Name	& R.A. (J2000)	& Declination (J2000)	& $S_{5.0}$ (Jy)	& $S_{30}$ (Jy)	& $\Delta S_{30}$ (Jy) & $\alpha^{5}_{1.4}$ & $\alpha^{30}_{5}$	& Morph.	\\
	(1)	& (2)	& (3)	& (4)	& (5)	& (6)	& (7)	& (8)	& (9)	\\
	\hline
	\endhead
	\hline
	\endfoot
	0003+380	& 00 05 57.2	& 38 20 15	& 0.549	& 0.654	& 0.037	& -0.07	& 0.10	& cj	\\
	0010+405	& 00 13 31.1	& 40 51 37	& 1.040	& 0.775	& 0.032	& -0.43	& -0.16	& cj	\\
	0014+813	& 00 17 08.5	& 81 35 08	& 0.551$^a$	& 0.629	& 0.030	& -0.17$^a$	& 0.07	& cj	\\
	0016+731	& 00 19 45.8	& 73 27 30	& 1.712	& 1.612	& 0.067	& 0.54	& -0.03	& cj	\\
	0018+729	& 00 21 27.4	& 73 12 41	& 0.397	& 0.071	& 0.004	& -0.42	& -0.96	& cj	\\
	0022+390	& 00 25 26.2	& 39 19 35	& 0.663	& 0.282	& 0.020	& -0.11	& -0.48	& cj	\\
	0035+367	& 00 37 46.1	& 36 59 10	& 0.482	& 0.184	& 0.010	& -0.47	& -0.54	& cj	\\
	0035+413	& 00 38 24.8	& 41 37 06	& 1.114	& 0.395	& 0.020	& 0.73	& -0.58	& cj	\\
	0102+480	& 01 05 49.9	& 48 19 03	& 1.088	& 0.428	& 0.024	& 0.15	& -0.52	& cj/c	\\
	0108+388	& 01 11 37.3	& 39 06 28	& 1.321	& 0.200	& 0.010	& 0.92	& -1.05	& cso	\\
	0109+351	& 01 12 12.9	& 35 22 19	& 0.362	& 1.014	& 0.043	& 0.00	& 0.57	& cj	\\
	0110+495	& 01 13 27.0	& 49 48 24	& 0.710	& 0.570	& 0.032	& 0.30	& -0.12	& cj	\\
	0133+476	& 01 36 58.6	& 47 51 29	& 1.816	& 3.271	& 0.171	& 0.21	& 0.33	& cj	\\
	0145+386	& 01 48 24.4	& 38 54 05	& 0.370	& 0.229	& 0.017	& 0.18	& -0.27	& cj?	\\
	0151+474	& 01 54 56.3	& 47 43 26	& 0.505	& 0.500	& 0.026	& 0.28	& -0.01	& cj	\\
	0153+744	& 01 57 35.0	& 74 42 43	& 1.549	& 0.148	& 0.007	& -0.23	& -1.31	& cj	\\
	0205+722	& 02 09 51.8	& 72 29 26	& 0.560	& 0.715	& 0.037	& -0.32	& 0.14	& cj	\\
	0212+735	& 02 17 30.8	& 73 49 32	& 2.278	& 2.726	& 0.136	& -0.11	& 0.10	& cj	\\
	0218+357	& 02 21 05.5	& 35 56 13	& 1.498	& 0.846	& 0.043	& 0.02	& -0.32	& g	\\
	0219+428	& 02 22 39.6	& 43 02 07	& 0.806	& 0.801	& 0.043	& 0.00	& -0.00	& cj	\\
	0227+403	& 02 30 45.7	& 40 32 53	& 0.436	& 0.421	& 0.022	& -0.00	& -0.02	& cj	\\
	0248+430	& 02 51 34.5	& 43 15 15	& 1.414	& 0.503	& 0.025	& 0.41	& -0.58	& cj/t	\\
	0249+383	& 02 53 08.9	& 38 35 24	& 0.450	& 0.286	& 0.015	& -0.43	& -0.25	& cj	\\
	0251+393	& 02 54 42.6	& 39 31 34	& 0.408	& 0.321	& 0.019	& 0.25	& -0.13	& cj	\\
	0256+424	& 02 59 37.7	& 42 35 49	& 0.366	& 0.132	& 0.009	& -0.41	& -0.57	& cj	\\
	0307+380	& 03 10 49.9	& 38 14 53	& 0.760	& 0.686	& 0.034	& 1.52	& -0.06	& c	\\
	0309+411	& 03 13 02.0	& 41 20 01	& 0.516	& 1.019	& 0.042	& 0.08	& 0.38	& cj	\\
	0316+413	& 03 19 48.2	& 41 30 42	& 42.370	& 10.803	& 0.507	& 0.54	& -0.76	& cj	\\
	0340+362	& 03 43 29.0	& 36 22 12	& 0.376	& 0.232	& 0.012	& 0.23	& -0.27	& cj?	\\
	0344+405	& 03 47 60.0	& 40 43 58	& 0.478	& 0.099	& 0.006	& -0.47	& -0.88	& d	\\
	0346+800	& 03 54 46.1	& 80 09 28	& 0.396$^a$	& 0.272	& 0.015	& -0.25$^a$	& -0.21	& cj	\\
	0402+379	& 04 05 49.3	& 38 03 32	& 0.937	& 0.285	& 0.016	& -0.38	& -0.66	& cj	\\
	0424+670	& 04 29 06.0	& 67 10 16	& 0.362	& 0.080	& 0.004	& -0.47	& -0.84	& c?	\\
	0444+634	& 04 49 23.3	& 63 32 09	& 0.606	& 0.699	& 0.034	& 0.39	& 0.08	& cj	\\
	0454+844	& 05 08 42.4	& 84 32 04	& 1.398$^a$	& 0.266	& 0.015	& 0.40$^a$	& -0.93	& cj	\\
	0537+531	& 05 41 16.2	& 53 12 24	& 0.665	& 0.230	& 0.011	& 0.01	& -0.59	& cj	\\
	0546+726	& 05 52 53.0	& 72 40 45	& 0.401	& 0.074	& 0.005	& -0.16	& -0.94	& cj?	\\
	0554+580	& 05 59 13.4	& 58 04 03	& 0.906	& 0.496	& 0.024	& 0.70	& -0.34	& cj	\\
	0600+442	& 06 04 35.6	& 44 13 58	& 0.705	& 0.212	& 0.014	& -0.42	& -0.67	& cj	\\
	0602+673	& 06 07 52.7	& 67 20 55	& 0.657	& 0.897	& 0.049	& 0.62	& 0.17	& cj	\\
	0604+728	& 06 10 48.9	& 72 48 53	& 0.654	& 0.273	& 0.015	& -0.35	& -0.49	& cj	\\
	0609+607	& 06 14 23.9	& 60 46 21	& 1.059	& 0.375	& 0.021	& -0.00	& -0.58	& cj	\\
	0615+820	& 06 26 03.0	& 82 02 25	& 0.999$^a$	& 0.343	& 0.021	& -0.03$^a$	& -0.60	& c	\\
	0620+389	& 06 24 19.0	& 38 56 48	& 0.811	& 0.450	& 0.028	& -0.32	& -0.33	& cj	\\
	0621+446	& 06 25 18.3	& 44 40 01	& 0.369	& 0.211	& 0.013	& 0.63	& -0.31	& c	\\
	0627+532	& 06 31 34.7	& 53 11 27	& 0.485	& 0.104	& 0.006	& -0.40	& -0.86	& cj	\\
	0633+596	& 06 38 02.9	& 59 33 22	& 0.482	& 0.691	& 0.037	& 0.58	& 0.20	& cj	\\
	0633+734	& 06 39 22.0	& 73 24 58	& 0.748	& 0.787	& 0.032	& -0.30	& 0.03	& cj	\\
	0636+680	& 06 42 04.3	& 67 58 35	& 0.499	& 0.113	& 0.006	& 1.06	& -0.83	& c	\\
	0641+393	& 06 44 53.7	& 39 14 47	& 0.453	& 0.322	& 0.016	& 0.17	& -0.19	& cj	\\
	0642+449	& 06 46 32.0	& 44 51 16	& 1.191	& 2.420	& 0.100	& 0.55	& 0.40	& cj?	\\
	0646+600	& 06 50 31.3	& 60 01 44	& 0.920	& 0.470	& 0.024	& 0.58	& -0.37	& d	\\
	0650+371	& 06 53 58.3	& 37 05 40	& 0.977	& 0.619	& 0.036	& 0.40	& -0.25	& cj	\\
	0650+453	& 06 54 23.7	& 45 14 23	& 0.420	& 0.516	& 0.027	& -0.27	& 0.11	& cj	\\
	0651+410	& 06 55 10.0	& 41 00 10	& 0.425	& 0.284	& 0.015	& 0.27	& -0.22	& cj?	\\
	0700+470	& 07 04 09.6	& 47 00 56	& 0.443	& 0.120	& 0.007	& -0.49	& -0.73	& cj	\\
	0702+612	& 07 07 00.6	& 61 10 11	& 0.370	& 0.433	& 0.024	& 0.10	& 0.09	& cj	\\
	0707+476	& 07 10 46.1	& 47 32 11	& 0.906	& 0.739	& 0.038	& -0.06	& -0.11	& cj	\\
	0710+439	& 07 13 38.2	& 43 49 17	& 1.629	& 0.423	& 0.024	& -0.09	& -0.75	& cso	\\
	0711+356	& 07 14 24.8	& 35 34 39	& 0.901	& 0.171	& 0.011	& -0.36	& -0.93	& d	\\
	0714+457	& 07 17 51.9	& 45 38 03	& 0.480	& 0.995	& 0.052	& 0.13	& 0.41	& cj	\\
	0716+714	& 07 21 53.4	& 71 20 36	& 0.788	& 0.546	& 0.029	& -0.02	& -0.20	& cj	\\
	0718+793	& 07 26 11.7	& 79 11 31	& 0.631$^a$	& 0.382	& 0.022	& 0.03$^a$	& -0.28	& cj	\\
	0724+571	& 07 28 49.6	& 57 01 24	& 0.393	& 0.389	& 0.021	& -0.03	& -0.01	& cj	\\
	0727+409	& 07 30 51.3	& 40 49 50	& 0.468	& 0.093	& 0.005	& 0.10	& -0.90	& cj	\\
	0730+504	& 07 33 52.5	& 50 22 09	& 0.890	& 0.428	& 0.026	& 0.66	& -0.41	& cj	\\
	0731+479	& 07 35 02.3	& 47 50 08	& 0.533	& 0.256	& 0.015	& 0.17	& -0.41	& cj	\\
	0733+597	& 07 37 30.1	& 59 41 03	& 0.357	& 0.165	& 0.009	& -0.34	& -0.43	& cj	\\
	0738+491	& 07 42 02.8	& 49 00 15	& 0.352	& 0.319	& 0.019	& 1.05	& -0.05	& cj?	\\
	0740+768	& 07 47 14.6	& 76 39 17	& 0.592$^a$	& 0.336	& 0.020	& 0.87$^a$	& -0.32	& d?	\\
	0743+744	& 07 49 22.5	& 74 20 41	& 0.479	& 0.306	& 0.016	& 0.24	& -0.25	& cj	\\
	0746+483	& 07 50 20.4	& 48 14 53	& 0.860	& 0.422	& 0.030	& 0.21	& -0.40	& cj	\\
	0749+426	& 07 53 03.3	& 42 31 30	& 0.461	& 0.058	& 0.004	& -0.33	& -1.16	& cj/d?	\\
	0749+540	& 07 53 01.4	& 53 52 59	& 0.877	& 0.913	& 0.051	& 0.08	& 0.02	& cj	\\
	0800+618	& 08 05 18.2	& 61 44 23	& 0.981	& 0.539	& 0.029	& 0.25	& -0.33	& cj	\\
	0803+452	& 08 06 33.5	& 45 04 32	& 0.414	& 0.350	& 0.018	& 0.05	& -0.09	& cj	\\
	0804+499	& 08 08 39.7	& 49 50 36	& 1.222	& 1.599	& 0.113	& 0.25	& 0.15	& cj	\\
	0805+410	& 08 08 56.7	& 40 52 44	& 0.743	& 0.694	& 0.036	& 0.57	& -0.04	& cj	\\
	0806+573	& 08 11 00.6	& 57 14 12	& 0.405	& 0.418	& 0.023	& -0.04	& 0.02	& cj	\\
	0812+367	& 08 15 25.9	& 36 35 15	& 0.980	& 0.414	& 0.022	& -0.03	& -0.48	& cj	\\
	0814+425	& 08 18 16.0	& 42 22 45	& 1.891	& 0.595	& 0.036	& 0.20	& -0.65	& cj	\\
	0820+560	& 08 24 47.2	& 55 52 42	& 1.199	& 0.595	& 0.031	& -0.05	& -0.39	& cj	\\
	0821+394	& 08 24 55.5	& 39 16 41	& 1.012	& 1.179	& 0.055	& -0.24	& 0.09	& cj	\\
	0821+621	& 08 25 38.6	& 61 57 28	& 0.615	& 0.294	& 0.017	& -0.05	& -0.41	& d	\\
	0824+355	& 08 27 38.6	& 35 25 05	& 0.746	& 0.294	& 0.016	& -0.12	& -0.52	& cj	\\
	0831+557	& 08 34 54.9	& 55 34 21	& 5.780	& 0.633	& 0.034	& -0.23	& -1.23	& cj	\\
	0833+416	& 08 36 36.9	& 41 25 54	& 0.385	& 0.306	& 0.016	& -0.08	& -0.13	& cj	\\
	0833+585	& 08 37 22.4	& 58 25 01	& 0.669	& 0.832	& 0.036	& 0.09	& 0.12	& cj	\\
	0836+710	& 08 41 24.4	& 70 53 42	& 2.423	& 1.309	& 0.064	& -0.44	& -0.34	& cj	\\
	0843+575	& 08 47 28.1	& 57 23 38	& 0.384	& 0.043	& 0.004	& 0.24	& -1.22	& cj	\\
	0847+379	& 08 50 24.7	& 37 47 09	& 0.382	& 0.287	& 0.016	& -0.37	& -0.16	& cj	\\
	0850+581	& 08 54 42.0	& 57 57 29	& 1.187	& 0.299	& 0.017	& -0.14	& -0.77	& cj	\\
	0859+470	& 09 03 04.0	& 46 51 04	& 1.285	& 0.823	& 0.044	& -0.45	& -0.25	& cj	\\
	0859+681	& 09 03 53.2	& 67 57 22	& 0.751	& 0.272	& 0.016	& 0.19	& -0.57	& cj	\\
	0900+520	& 09 03 58.6	& 51 51 00	& 0.395	& 0.182	& 0.010	& 0.16	& -0.43	& cj?	\\
	0902+490	& 09 05 27.5	& 48 50 49	& 0.547	& 0.096	& 0.005	& -0.12	& -0.97	& cj?	\\
	0917+449	& 09 20 58.5	& 44 41 53	& 1.033	& 1.704	& 0.101	& 0.22	& 0.28	& cj	\\
	0917+624	& 09 21 36.2	& 62 15 52	& 1.322	& 0.813	& 0.040	& 0.06	& -0.27	& d	\\
	0923+392	& 09 27 03.0	& 39 02 20	& 7.480	& 5.726	& 0.388	& 0.80	& -0.15	& cj	\\
	0925+504	& 09 29 15.4	& 50 13 35	& 0.558	& 0.469	& 0.027	& 0.58	& -0.10	& cj	\\
	0927+352	& 09 30 55.3	& 35 03 37	& 0.383	& 0.294	& 0.018	& -0.08	& -0.15	& cj	\\
	0929+533	& 09 32 41.2	& 53 06 33	& 0.384	& 0.141	& 0.008	& -0.26	& -0.56	& cj	\\
	0930+493	& 09 34 15.8	& 49 08 21	& 0.574	& 0.161	& 0.009	& -0.19	& -0.71	& cj	\\
	0942+468	& 09 45 42.1	& 46 36 50	& 0.354	& 0.192	& 0.010	& 0.19	& -0.34	& cj	\\
	0945+408	& 09 48 55.3	& 40 39 44	& 1.592	& 1.018	& 0.076	& 0.05	& -0.25	& cj	\\
	0945+664	& 09 49 12.2	& 66 14 59	& 1.407	& 0.270	& 0.016	& -0.36	& -0.92	& mso$^b$	\\
	0949+354	& 09 52 32.0	& 35 12 52	& 0.403	& 0.216	& 0.013	& 0.12	& -0.35	& cj	\\
	0950+748	& 09 54 47.4	& 74 35 57	& 0.738	& 0.106	& 0.007	& -0.37	& -1.08	& cj	\\
	0954+556	& 09 57 38.2	& 55 22 57	& 2.270	& 1.016	& 0.076	& -0.22	& -0.45	& mso$^b$	\\
	0954+658	& 09 58 47.2	& 65 33 54	& 1.417	& 1.002	& 0.075	& 0.61	& -0.19	& cj	\\
	0955+476	& 09 58 19.7	& 47 25 07	& 0.834	& 1.389	& 0.059	& 0.15	& 0.28	& cj	\\
	1003+830	& 10 10 15.8	& 82 50 14	& 0.716$^a$	& 0.422	& 0.034	& -0.14$^a$	& -0.30	& cj	\\
	1010+350	& 10 13 49.6	& 34 45 50	& 0.597	& 0.420	& 0.025	& 0.28	& -0.20	& cj	\\
	1014+615	& 10 17 25.9	& 61 16 27	& 0.631	& 0.302	& 0.016	& 0.44	& -0.41	& cj	\\
	1015+359	& 10 18 11.0	& 35 42 39	& 0.587	& 1.007	& 0.058	& -0.34	& 0.30	& cj	\\
	1020+400	& 10 23 11.6	& 39 48 15	& 0.785	& 0.571	& 0.029	& -0.31	& -0.18	& cj	\\
	1030+398	& 10 33 22.1	& 39 35 51	& 0.645	& 0.128	& 0.007	& 0.42	& -0.90	& cj	\\
	1030+415	& 10 33 03.7	& 41 16 06	& 0.485	& 0.617	& 0.036	& -0.37	& 0.13	& cj	\\
	1030+611	& 10 33 51.4	& 60 51 07	& 0.579	& 0.490	& 0.026	& -0.22	& -0.09	& cj	\\
	1031+567	& 10 35 07.0	& 56 28 46	& 1.200	& 0.184	& 0.010	& -0.30	& -1.05	& cso	\\
	1038+528	& 10 41 46.8	& 52 33 28	& 0.709	& 0.393	& 0.020	& -0.00	& -0.33	& cj	\\
	1039+811	& 10 44 23.1	& 80 54 39	& 1.144$^a$	& 0.864	& 0.064	& 0.40$^a$	& -0.16	& cj	\\
	1041+536	& 10 44 10.7	& 53 22 20	& 0.481	& 0.370	& 0.019	& -0.10	& -0.15	& cj	\\
	1044+719	& 10 48 27.6	& 71 43 35	& 2.410	& 2.480	& 0.109	& 1.06	& 0.02	& c	\\
	1053+704	& 10 56 53.6	& 70 11 45	& 0.675	& 1.073	& 0.060	& 0.08	& 0.26	& cj	\\
	1053+815	& 10 58 11.5	& 81 14 32	& 0.770$^a$	& 0.643	& 0.047	& -0.36$^a$	& -0.10	& cj	\\
	1058+629	& 11 01 53.4	& 62 41 50	& 0.700	& 0.105	& 0.006	& 0.12	& -1.06	& cj	\\
	1058+726	& 11 01 48.8	& 72 25 37	& 0.953	& 0.493	& 0.030	& -0.33	& -0.37	& cj	\\
	1101+384	& 11 04 27.3	& 38 12 31	& 0.722	& 0.393	& 0.020	& -0.11	& -0.34	& cj	\\
	1105+437	& 11 08 23.5	& 43 30 53	& 0.375	& 0.382	& 0.023	& 0.27	& 0.01	& cj	\\
	1106+380	& 11 09 28.9	& 37 44 30	& 0.867	& 0.240	& 0.013	& -0.38	& -0.72	& cj	\\
	1107+607	& 11 10 13.1	& 60 28 42	& 0.404	& 0.053	& 0.004	& 0.12	& -1.13	& cj	\\
	1124+455	& 11 26 57.7	& 45 16 06	& 0.355	& 0.060	& 0.004	& -0.26	& -0.99	& cj	\\
	1124+571	& 11 27 40.1	& 56 50 14	& 0.597	& 0.237	& 0.014	& -0.20	& -0.52	& cj	\\
	1125+596	& 11 28 13.3	& 59 25 14	& 0.393	& 0.605	& 0.034	& 0.06	& 0.24	& c	\\
	1128+385	& 11 30 53.3	& 38 15 18	& 0.746	& 1.094	& 0.055	& -0.18	& 0.21	& cj	\\
	1143+590	& 11 46 26.9	& 58 48 34	& 0.674	& 0.422	& 0.025	& 0.70	& -0.26	& cj	\\
	1144+352	& 11 47 22.1	& 35 01 07	& 0.663	& 0.182	& 0.010	& -0.04	& -0.72	& cj	\\
	1144+402	& 11 46 58.3	& 39 58 34	& 0.739	& 1.129	& 0.047	& -0.17	& 0.24	& cj	\\
	1144+542	& 11 46 44.2	& 53 56 43	& 0.484	& 0.391	& 0.024	& 0.14	& -0.12	& cj	\\
	1146+596	& 11 48 50.4	& 59 24 56	& 0.627	& 0.362	& 0.022	& 0.32	& -0.31	& cj/cso	\\
	1150+812	& 11 53 12.5	& 80 58 29	& 1.181$^a$	& 1.522	& 0.091	& -0.09$^a$	& 0.14	& cj	\\
	1151+408	& 11 53 54.7	& 40 36 52	& 0.380	& 0.246	& 0.013	& -0.48	& -0.24	& cj?	\\
	1155+486	& 11 58 26.8	& 48 25 16	& 0.445	& 0.214	& 0.013	& -0.07	& -0.41	& cj	\\
	1205+544	& 12 08 27.5	& 54 13 19	& 0.397	& 0.091	& 0.006	& -0.12	& -0.82	& d?	\\
	1206+415	& 12 09 22.8	& 41 19 41	& 0.515	& 0.197	& 0.010	& 0.37	& -0.54	& c	\\
	1213+350	& 12 15 55.6	& 34 48 15	& 1.152	& 0.540	& 0.030	& -0.32	& -0.42	& cj	\\
	1216+487	& 12 19 06.4	& 48 29 56	& 0.680	& 0.670	& 0.040	& -0.18	& -0.01	& cj	\\
	1218+444	& 12 21 27.0	& 44 11 29	& 0.478	& 0.206	& 0.013	& -0.18	& -0.47	& cj	\\
	1221+809	& 12 23 40.5	& 80 40 04	& 0.518$^a$	& 0.392	& 0.028	& 0.43$^a$	& -0.16	& cj	\\
	1223+395	& 12 25 50.6	& 39 14 22	& 0.438	& 0.154	& 0.009	& -0.16	& -0.58	& cj	\\
	1226+373	& 12 28 47.4	& 37 06 12	& 0.953	& 0.168	& 0.009	& 1.26	& -0.97	& cj	\\
	1239+376	& 12 42 09.8	& 37 20 05	& 0.446	& 0.203	& 0.012	& -0.16	& -0.44	& cj	\\
	1240+381	& 12 42 51.4	& 37 51 00	& 0.768	& 0.375	& 0.022	& 0.59	& -0.40	& cj	\\
	1246+586	& 12 48 18.8	& 58 20 28	& 0.414	& 0.123	& 0.008	& 0.43	& -0.68	& cj	\\
	1250+532	& 12 53 11.9	& 53 01 11	& 0.396	& 0.380	& 0.023	& -0.24	& -0.02	& cj	\\
	1254+571	& 12 56 14.2	& 56 52 25	& 0.419	& 0.129	& 0.009	& 0.29	& -0.66	& cj?	\\
	1258+507	& 13 00 41.2	& 50 29 36	& 0.391	& 0.280	& 0.015	& -0.23	& -0.19	& cj	\\
	1300+580	& 13 02 52.5	& 57 48 37	& 0.758	& 0.759	& 0.047	& 0.72	& 0.00	& cj	\\
	1305+804	& 13 06 05.7	& 80 08 20	& 0.375$^a$	& 0.108	& 0.012	& -0.47$^a$	& -0.69	& cj	\\
	1306+360	& 13 08 23.7	& 35 46 37	& 0.437	& 0.510	& 0.029	& 0.54	& 0.09	& cj	\\
	1307+562	& 13 09 09.8	& 55 57 38	& 0.416	& 0.279	& 0.018	& 0.27	& -0.22	& cj	\\
	1308+471	& 13 10 53.6	& 46 53 52	& 0.393	& 0.281	& 0.017	& 1.08	& -0.19	& cj?	\\
	1309+555	& 13 11 03.2	& 55 13 54	& 0.677	& 0.278	& 0.017	& 0.91	& -0.50	& cj	\\
	1312+533	& 13 14 43.8	& 53 06 27	& 0.433	& 0.066	& 0.004	& 0.49	& -1.05	& cj	\\
	1321+410	& 13 24 12.1	& 40 48 11	& 0.413	& 0.057	& 0.004	& 0.11	& -1.11	& d	\\
	1322+835	& 13 21 45.6	& 83 16 13	& 0.506$^a$	& 0.194	& 0.015	& 0.23$^a$	& -0.54	& cj	\\
	1323+800	& 13 23 51.6	& 79 42 51	& 0.458$^a$	& 0.307	& 0.019	& 0.22$^a$	& -0.22	& cj	\\
	1325+436	& 13 27 21.0	& 43 26 27	& 0.533	& 0.194	& 0.011	& -0.22	& -0.56	& cj	\\
	1333+459	& 13 35 22.0	& 45 42 38	& 0.598	& 0.230	& 0.013	& 0.51	& -0.53	& cj?	\\
	1333+589	& 13 35 25.9	& 58 44 00	& 0.820	& 0.200	& 0.011	& -0.07	& -0.79	& d	\\
	1335+552	& 13 37 49.6	& 55 01 02	& 0.811	& 0.378	& 0.023	& 0.10	& -0.43	& cj	\\
	1337+637	& 13 39 23.8	& 63 28 58	& 0.431	& 0.080	& 0.005	& -0.12	& -0.94	& cj	\\
	1342+663	& 13 44 08.7	& 66 06 11	& 0.510	& 0.300	& 0.017	& -0.43	& -0.30	& cj	\\
	1347+539	& 13 49 34.7	& 53 41 17	& 0.635	& 0.438	& 0.024	& -0.46	& -0.21	& cj	\\
	1355+441	& 13 57 40.7	& 43 53 59	& 0.464	& 0.101	& 0.007	& -0.33	& -0.85	& cj	\\
	1356+478	& 13 58 40.7	& 47 37 58	& 0.428	& 0.070	& 0.006	& -0.26	& -1.01	& d/cso?	\\
	1357+769	& 13 57 55.4	& 76 43 21	& 0.844$^a$	& 0.698	& 0.043	& 0.62$^a$	& -0.11	& cj	\\
	1413+373	& 14 15 28.5	& 37 06 21	& 0.383	& 0.060	& 0.004	& 0.04	& -1.03	& cj	\\
	1415+463	& 14 17 08.2	& 46 07 05	& 0.904	& 0.489	& 0.026	& -0.09	& -0.34	& cj	\\
	1417+385	& 14 19 46.6	& 38 21 48	& 0.871	& 0.543	& 0.032	& 0.16	& -0.26	& cj	\\
	1418+546	& 14 19 46.6	& 54 23 14	& 1.707	& 0.781	& 0.033	& 0.07	& -0.44	& cj	\\
	1421+482	& 14 23 06.2	& 48 02 10	& 0.536	& 0.249	& 0.014	& 0.31	& -0.43	& cj	\\
	1424+366	& 14 26 37.1	& 36 25 09	& 0.429	& 0.278	& 0.015	& 0.62	& -0.24	& cj	\\
	1427+543	& 14 29 21.9	& 54 06 11	& 0.718	& 0.178	& 0.009	& -0.18	& -0.78	& cj	\\
	1432+422	& 14 34 05.7	& 42 03 16	& 0.353	& 0.109	& 0.007	& 0.19	& -0.66	& cj	\\
	1435+638	& 14 36 45.8	& 63 36 37	& 0.795	& 0.541	& 0.032	& -0.44	& -0.21	& cj	\\
	1438+385	& 14 40 22.3	& 38 20 13	& 0.944	& 0.321	& 0.018	& -0.06	& -0.60	& cj	\\
	1442+637	& 14 43 58.6	& 63 32 26	& 0.456	& 0.088	& 0.006	& -0.32	& -0.92	& cj	\\
	1448+762	& 14 48 28.8	& 76 01 11	& 0.683$^a$	& 0.501	& 0.028	& 0.33$^a$	& -0.17	& cj	\\
	1456+375	& 14 58 44.8	& 37 20 21	& 0.591	& 0.151	& 0.008	& 0.45	& -0.76	& cj	\\
	1459+480	& 15 00 48.7	& 47 51 15	& 0.489	& 0.282	& 0.016	& 0.16	& -0.31	& cj	\\
	1504+377	& 15 06 09.5	& 37 30 51	& 1.003	& 0.555	& 0.034	& -0.13	& -0.33	& cj	\\
	1505+428	& 15 06 53.0	& 42 39 23	& 0.404	& 0.547	& 0.030	& -0.06	& 0.17	& cj	\\
	1526+670	& 15 26 42.9	& 66 50 54	& 0.417	& 0.078	& 0.005	& -0.02	& -0.94	& cj	\\
	1531+722	& 15 31 33.6	& 72 06 41	& 0.452	& 0.161	& 0.009	& -0.30	& -0.58	& cj	\\
	1534+501	& 15 35 52.0	& 49 57 39	& 0.359	& 0.169	& 0.010	& 0.35	& -0.42	& c	\\
	1543+480	& 15 45 08.5	& 47 51 54	& 0.441	& 0.078	& 0.006	& -0.32	& -0.97	& d/cso	\\
	1543+517	& 15 45 02.8	& 51 35 00	& 0.544	& 0.093	& 0.005	& 0.09	& -0.99	& cj	\\
	1545+497	& 15 47 21.1	& 49 37 05	& 0.549	& 0.163	& 0.010	& -0.42	& -0.68	& c	\\
	1547+507	& 15 49 17.5	& 50 38 05	& 0.724	& 0.738	& 0.039	& 0.06	& 0.01	& cj	\\
	1550+582	& 15 51 58.2	& 58 06 44	& 0.367	& 0.254	& 0.014	& 0.38	& -0.21	& cj	\\
	1619+491	& 16 20 31.2	& 49 01 53	& 0.469	& 0.214	& 0.012	& 0.00	& -0.44	& cj	\\
	1622+665	& 16 23 04.5	& 66 24 01	& 0.520	& 0.122	& 0.007	& 0.74	& -0.81	& cj?	\\
	1623+578	& 16 24 24.8	& 57 41 16	& 0.590	& 0.424	& 0.025	& 0.13	& -0.18	& cj	\\
	1624+416	& 16 25 57.7	& 41 34 40	& 1.362	& 0.521	& 0.030	& -0.17	& -0.54	& cj	\\
	1629+495	& 16 31 16.5	& 49 27 39	& 0.394	& 0.458	& 0.026	& 0.16	& 0.08	& cj	\\
	1633+382	& 16 35 15.5	& 38 08 04	& 3.189	& 2.976	& 0.173	& 0.41	& -0.04	& cj	\\
	1636+473	& 16 37 45.1	& 47 17 33	& 1.330	& 1.031	& 0.053	& 0.27	& -0.14	& cj	\\
	1637+574	& 16 38 13.5	& 57 20 23	& 1.807	& 1.317	& 0.071	& 0.30	& -0.18	& c	\\
	1638+398	& 16 40 29.6	& 39 46 46	& 1.285	& 0.612	& 0.025	& 0.53	& -0.41	& cj?	\\
	1638+540	& 16 39 39.8	& 53 57 47	& 0.369	& 0.172	& 0.011	& 0.14	& -0.43	& cj	\\
	1641+399	& 16 42 58.8	& 39 48 36	& 8.363	& 4.929	& 0.287	& 0.05	& -0.30	& cj	\\
	1642+690	& 16 42 07.8	& 68 56 39	& 1.516	& 2.746	& 0.128	& 0.00	& 0.33	& d/cj?	\\
	1645+410	& 16 46 56.9	& 40 59 17	& 0.388	& 0.185	& 0.009	& 0.16	& -0.41	& cj	\\
	1645+635	& 16 45 58.6	& 63 30 10	& 0.444	& 0.410	& 0.023	& 0.31	& -0.04	& cj	\\
	1652+398	& 16 53 52.2	& 39 45 36	& 1.371	& 1.219	& 0.071	& -0.04	& -0.07	& cj	\\
	1656+477	& 16 58 02.8	& 47 37 49	& 1.420	& 0.582	& 0.031	& 0.47	& -0.50	& cj	\\
	1656+482	& 16 57 46.9	& 48 08 33	& 0.847	& 0.697	& 0.045	& -0.12	& -0.11	& cj	\\
	1656+571	& 16 57 20.7	& 57 05 53	& 0.844	& 0.487	& 0.027	& 0.04	& -0.31	& cj	\\
	1700+685	& 17 00 09.3	& 68 30 06	& 0.435	& 0.338	& 0.017	& 0.29	& -0.14	& cj	\\
	1716+686	& 17 16 13.9	& 68 36 38	& 0.988	& 0.536	& 0.025	& 0.69	& -0.34	& cj	\\
	1719+357	& 17 21 09.5	& 35 42 16	& 0.874	& 0.210	& 0.011	& 0.03	& -0.80	& cj	\\
	1722+401	& 17 24 05.4	& 40 04 36	& 0.532	& 0.619	& 0.037	& -0.03	& 0.08	& cj	\\
	1726+455	& 17 27 27.7	& 45 30 39	& 1.066	& 0.715	& 0.043	& 0.72	& -0.22	& cj	\\
	1732+389	& 17 34 20.6	& 38 57 51	& 0.561	& 1.105	& 0.075	& -0.26	& 0.38	& cj	\\
	1734+508	& 17 35 49.0	& 50 49 11	& 0.798	& 0.420	& 0.028	& 0.06	& -0.36	& d	\\
	1738+476	& 17 39 57.1	& 47 37 58	& 0.789	& 0.621	& 0.038	& -0.04	& -0.13	& cj	\\
	1738+499	& 17 39 27.4	& 49 55 03	& 0.478	& 0.427	& 0.026	& -0.13	& -0.06	& cj	\\
	1739+522	& 17 40 37.0	& 52 11 43	& 1.133	& 0.826	& 0.045	& -0.44	& -0.18	& cj	\\
	1744+557	& 17 44 56.6	& 55 42 17	& 0.599	& 0.301	& 0.017	& -0.17	& -0.38	& cj	\\
	1745+624	& 17 46 14.0	& 62 26 54	& 0.580	& 0.205	& 0.011	& -0.22	& -0.58	& cj	\\
	1746+470	& 17 47 26.6	& 46 58 50	& 0.634	& 0.194	& 0.010	& 0.31	& -0.66	& cj?	\\
	1747+433	& 17 49 00.4	& 43 21 51	& 0.367	& 0.423	& 0.026	& 0.06	& 0.08	& cj	\\
	1749+701	& 17 48 32.8	& 70 05 50	& 0.728	& 0.430	& 0.023	& -0.46	& -0.29	& cj	\\
	1751+441	& 17 53 22.6	& 44 09 45	& 0.998	& 0.551	& 0.029	& 0.19	& -0.33	& cj	\\
	1755+578	& 17 56 03.6	& 57 48 47	& 0.455	& 0.062	& 0.004	& -0.37	& -1.11	& cj/cso?	\\
	1758+388	& 18 00 24.8	& 38 48 30	& 0.722	& 0.876	& 0.037	& 0.27	& 0.11	& cj	\\
	1800+440	& 18 01 32.3	& 44 04 21	& 1.148	& 0.736	& 0.042	& 0.21	& -0.25	& cj	\\
	1803+784	& 18 00 45.7	& 78 28 04	& 2.633$^a$	& 1.211	& 0.063	& 0.27$^a$	& -0.43	& cj	\\
	1807+698	& 18 06 50.7	& 69 49 28	& 2.189	& 1.493	& 0.080	& -0.03	& -0.21	& cj	\\
	1809+568	& 18 10 03.3	& 56 49 22	& 0.576	& 0.269	& 0.013	& 0.01	& -0.42	& cj	\\
	1812+412	& 18 14 22.7	& 41 13 05	& 0.534	& 0.266	& 0.018	& -0.15	& -0.39	& cj	\\
	1818+356	& 18 20 42.1	& 35 40 39	& 0.573	& 0.195	& 0.012	& -0.43	& -0.60	& cj	\\
	1823+568	& 18 24 07.1	& 56 51 01	& 1.135	& 1.642	& 0.100	& -0.21	& 0.21	& cj	\\
	1826+796	& 18 23 14.1	& 79 38 49	& 0.577$^a$	& 0.169	& 0.008	& 0.41$^a$	& -0.69	& cso	\\
	1828+399	& 18 29 56.5	& 39 57 34	& 0.353	& 0.266	& 0.014	& 0.80	& -0.16	& cj?	\\
	1834+612	& 18 35 19.7	& 61 19 40	& 0.590	& 0.202	& 0.011	& 0.22	& -0.60	& cj	\\
	1839+389	& 18 40 57.2	& 39 00 45	& 0.476	& 0.171	& 0.010	& 0.26	& -0.57	& cj?	\\
	1842+681	& 18 42 33.6	& 68 09 25	& 0.936	& 0.986	& 0.055	& 0.34	& 0.03	& cj	\\
	1843+356	& 18 45 35.1	& 35 41 16	& 0.794	& 0.111	& 0.007	& -0.21	& -1.10	& cj	\\
	1849+670	& 18 49 16.1	& 67 05 41	& 0.992	& 0.575	& 0.029	& 0.08	& -0.30	& cj	\\
	1850+402	& 18 52 30.4	& 40 19 06	& 0.535	& 0.382	& 0.025	& -0.02	& -0.19	& cj	\\
	1851+488	& 18 52 28.5	& 48 55 47	& 0.351	& 0.196	& 0.011	& 0.15	& -0.33	& c	\\
	1856+737	& 18 54 57.3	& 73 51 19	& 0.546	& 0.530	& 0.027	& -0.02	& -0.02	& cj	\\
	1908+484	& 19 09 46.6	& 48 34 31	& 0.423	& 0.151	& 0.008	& -0.24	& -0.57	& c	\\
	1910+375	& 19 12 25.1	& 37 40 36	& 0.402	& 0.282	& 0.016	& -0.18	& -0.20	& cj	\\
	1924+507	& 19 26 06.3	& 50 52 57	& 0.354	& 0.213	& 0.011	& -0.48	& -0.28	& cj	\\
	1926+611	& 19 27 30.4	& 61 17 32	& 0.618	& 0.837	& 0.044	& -0.06	& 0.17	& cj	\\
	1928+738	& 19 27 48.5	& 73 58 01	& 3.561	& 2.961	& 0.159	& -0.07	& -0.10	& cj	\\
	1936+714	& 19 36 03.6	& 71 31 31	& 0.391	& 0.235	& 0.012	& -0.36	& -0.28	& cj?	\\
	1943+546	& 19 44 31.5	& 54 48 07	& 0.938	& 0.225	& 0.011	& -0.44	& -0.80	& d	\\
	1946+708	& 19 45 53.5	& 70 55 48	& 0.645	& 0.217	& 0.011	& -0.28	& -0.61	& cso	\\
	1950+573	& 19 51 07.0	& 57 27 17	& 0.476	& 0.226	& 0.011	& -0.14	& -0.42	& cj	\\
	1954+513	& 19 55 42.7	& 51 31 48	& 1.610	& 0.891	& 0.047	& -0.01	& -0.33	& cj	\\
	2005+642	& 20 06 17.7	& 64 24 45	& 0.739	& 0.234	& 0.012	& 1.14	& -0.64	& cj	\\
	2007+659	& 20 07 28.8	& 66 07 22	& 0.756	& 0.463	& 0.024	& -0.24	& -0.27	& cj	\\
	2007+777	& 20 05 31.0	& 77 52 43	& 1.279$^a$	& 1.117	& 0.046	& 0.73$^a$	& -0.08	& cj	\\
	2010+723	& 20 09 52.3	& 72 29 19	& 0.910	& 0.916	& 0.045	& -0.18	& 0.00	& cj	\\
	2017+745	& 20 17 13.1	& 74 40 48	& 0.500	& 0.309	& 0.016	& 0.05	& -0.27	& cj	\\
	2021+614	& 20 22 06.7	& 61 36 58	& 2.743	& 1.433	& 0.060	& 0.20	& -0.36	& d?	\\
	2023+760	& 20 22 35.6	& 76 11 26	& 0.426$^a$	& 0.479	& 0.025	& -0.13$^a$	& 0.07	& cj	\\
	2054+611	& 20 55 38.8	& 61 22 00	& 0.414	& 0.262	& 0.014	& -0.02	& -0.26	& cj	\\
	2116+818	& 21 14 01.2	& 82 04 48	& 0.376$^a$	& 0.224	& 0.012	& -0.22$^a$	& -0.29	& cj	\\
	2136+824	& 21 33 34.1	& 82 39 06	& 0.509$^a$	& 0.161	& 0.009	& -0.48$^a$	& -0.64	& cj	\\
	2138+389	& 21 40 16.9	& 39 11 44	& 0.502	& 0.226	& 0.012	& -0.22	& -0.45	& cj	\\
	2200+420	& 22 02 43.3	& 42 16 39	& 3.593	& 5.570	& 0.301	& -0.21	& 0.24	& cj	\\
	2214+350	& 22 16 20.0	& 35 18 14	& 0.477	& 0.319	& 0.018	& -0.06	& -0.22	& cj?	\\
	2229+695	& 22 30 36.5	& 69 46 28	& 1.365	& 0.591	& 0.031	& 0.80	& -0.47	& cj	\\
	2235+731	& 22 36 38.6	& 73 22 52	& 0.424	& 0.165	& 0.009	& 0.27	& -0.53	& cj	\\
	2238+410	& 22 41 07.2	& 41 20 11	& 0.677	& 0.226	& 0.012	& 0.12	& -0.61	& cj	\\
	2253+417	& 22 55 36.7	& 42 02 52	& 1.120	& 0.577	& 0.031	& -0.18	& -0.37	& cj	\\
	2255+416	& 22 57 22.1	& 41 54 16	& 1.111	& 0.216	& 0.012	& -0.41	& -0.91	& cj	\\
	2259+371	& 23 01 27.7	& 37 26 49	& 0.406	& 0.513	& 0.026	& -0.30	& 0.13	& cj	\\
	2309+454	& 23 11 47.4	& 45 43 56	& 0.597	& 0.401	& 0.020	& 0.53	& -0.22	& cj	\\
	2310+385	& 23 12 58.8	& 38 47 42	& 0.484	& 0.166	& 0.009	& -0.28	& -0.60	& d	\\
	2319+444	& 23 22 20.4	& 44 45 42	& 0.366	& 0.282	& 0.015	& 0.14	& -0.15	& cj	\\
	2346+385	& 23 49 20.8	& 38 49 17	& 0.640	& 0.535	& 0.029	& 0.54	& -0.10	& cj	\\
	2351+456	& 23 54 21.7	& 45 53 04	& 1.145	& 1.223	& 0.062	& -0.37	& 0.04	& cj	\\
	2352+495	& 23 55 09.5	& 49 50 08	& 1.552	& 0.328	& 0.020	& -0.32	& -0.87	& cso	\\
	2353+816	& 23 56 22.8	& 81 52 52	& 0.476$^a$	& 0.526	& 0.025	& 0.13$^a$	& 0.06	& cj	\\
	2356+385	& 23 59 33.2	& 38 50 42	& 0.449	& 0.188	& 0.010	& -0.28	& -0.49	& cj?	\\
	2356+390	& 23 58 59.9	& 39 22 28	& 0.371	& 0.414	& 0.020	& -0.11	& 0.06	& cj	\\
	\end{longtable}
	$^a$ These spectral indicies are calculated between 2.727~GHz and 5~GHz using data from \cite{1981AJ.....86..854K}.
	$^b$ These identifications were made by \cite{2005A&A...434..449R}.
} 

\end{document}